\documentclass[
    aps,
    prfluids,
    onecolumn,
    notitlepage,
    superscriptaddress,
    longbibliography,
    floatfix]{revtex4-2}

\usepackage{graphicx}
\usepackage{dcolumn}
\usepackage{bm}
\usepackage{amsmath,amssymb}
\usepackage[colorlinks=true,allcolors=blue]{hyperref}
\newcommand{\markblue}[1]{{\leavevmode\color{black}{#1}}}

\begin{document}

\title{Hydrodynamic Resistance on Oscillating Planar Interfacial Bodies}

\author{Ian Ho}
\affiliation{Brown University, Center for Fluid Mechanics and School of Engineering, 184 Hope St., Providence, RI 02912, USA}
\affiliation{Department of Bioengineering, Stanford University, Stanford CA 94305, USA}

\author{Ajay Harishankar Kumar}
\affiliation{Brown University, Center for Fluid Mechanics and School of Engineering, 184 Hope St., Providence, RI 02912, USA}
\affiliation{Davidson School of Chemical Engineering, Purdue University, West Lafayette, IN 47907, USA}

\author{Daniel M. Harris}
\email{daniel\_harris3@brown.edu}
\affiliation{Brown University, Center for Fluid Mechanics and School of Engineering, 184 Hope St., Providence, RI 02912, USA}

\date{\today}

\begin{abstract}
We study the unsteady dynamics of floating planar bodies undergoing lateral oscillations along an air--water interface. Scaling arguments indicate that \markblue{when the viscous penetration depth and oscillation
amplitude are both small compared to the body size}, the flow beneath the body can be approximated by an oscillatory Stokes boundary layer, yielding a leading-order description of the hydrodynamic resistance. Using magnetic actuation, we drive the interfacial bodies harmonically and measure the amplitude response and phase lag in steady state over a range of frequencies, masses, sizes, and shapes. This frequency-response framework enables direct extraction of effective added mass and damping coefficients, which we find to be consistent with oscillatory boundary-layer theory in the limit of small interfacial deformation. The transient behavior during startup is also shown to be accurately predicted by a history integral that captures the development of the oscillatory boundary layer beneath the body.  This work also establishes a simple experimental platform for quantifying unsteady hydrodynamic forces at fluid interfaces.
\end{abstract}

\maketitle

\section{Introduction}
At a fluid interface, an object in motion does not simply remain in uniform motion; it is acted on and slowed by hydrodynamic forces. Depending on the geometry, scale, and operating regime, this resistance may arise from vortex-induced drag, wave-induced drag, or boundary-layer skin friction. Beyond such dissipative resistances, an accelerating body also experiences \markblue{an added-mass force} associated with the inertia of the surrounding fluid that must be set in motion. This added-mass causes the body to behave as though it were heavier. \markblue{Here and throughout, we refer to this acceleration-proportional added-mass contribution as the reactive component of the hydrodynamic force \cite{leonard2001aspects}.} For ships, boats, and other large floating bodies, the relative importance of these dissipative and reactive mechanisms has been extensively studied \cite{faltinsen2005hydrodynamics,barratt1965wave,newman1977,morison1950,sarpkaya2010}. For bodies at capillary or centimeter scales, however, theoretical descriptions and experimental studies of unsteady hydrodynamic resistance are more limited. The associated hydrodynamic forces in this centimeter-scale regime are important both for understanding natural systems and for guiding the design of engineered ones, from organisms that locomote at or near the air--water interface \cite{bush2006walking,hu2010hydrodynamics,suter1997locomotion,mukundarajan2016surface,hsieh2004running,roh2019honeybees,jung2026ground}, to self-organizing interfacial active matter \cite{ho2023capillary,oza2023theoretical,harris2025propulsion,barotta2025synchronization,sungar2025synchronization,izri2014self,bormashenko2015self}, to self-propelled interfacial robots \cite{rhee2022surferbot,benham2024wave,o2026achieving,tarr2024probing,hartmann2025highly,chen2018controllable,song2007surface}. Across these settings, the hydrodynamic forces are shaped not only by the body's geometry and motion, but also by the unique physics of the interface itself.

At an interface, partial immersion and wetting \cite{hunt2023drag}, attached interfaces \cite{dorr2016drag,zhou2022drag,petkov1995measurement,ally2010magnetophoretic}, and contact-line dynamics or meniscus deformation \cite{loudet2020drag,le2011wave} can all modify the near-body flow and affect drag and added mass. For spherical particles straddling or partially immersed at a fluid interface, an extensive body of theoretical, numerical, and experimental work has characterized corrections to the Stokes drag arising from the presence of the interface, parameterized by the contact angle, viscosity ratio, degree of immersion, and interfacial deformation~\cite{petkov1995measurement,danov1995influence,danov2000viscous,pozrikidis2007particle,ally2010magnetophoretic,dani2015hydrodynamics,dorr2016drag,zhou2022drag,hunt2023drag,kamoliddinov2021hydrodynamic,loudet2020drag}. For freely decelerating disks on a water surface, the hydrodynamic resistance has been modeled by a quasi-static Blasius boundary layer, assuming a flat interface~\cite{pucci2019friction}. In all these cases, however, the flow is treated as quasi-steady, with only the dissipative component of the hydrodynamic force resolved which acts solely to resist the instantaneous velocity of the object. Comparatively little is known about the unsteady counterpart, in which both \markblue{added-mass (reactive)} and \markblue{viscous-damping (dissipative)} components contribute, and even less for non-spherical geometries such as the planar bodies considered here.

A natural starting point for the unsteady regime is Stokes' second problem. Since the time of Prandtl \cite{anderson2005ludwig}, it has long been known that viscous stresses confined to thin fluid boundary layers can exert leading-order influence over the force on a moving body. In the unsteady setting, the canonical example is the oscillation of an infinite plate, which generates an oscillatory boundary layer with an analytical prediction for the viscous stress near the no-slip wall \cite{stokes1922mathematical,rayleigh1911lxxxii, schlichting2016boundary}. Related work has extended this problem to full analytical descriptions of transient start-up \cite{panton1968transient,liu2006note,erdogan2000note}, cylindrical geometries \cite{li1971stokes,rivero2019study,song2020viscous}, and the linear stability of oscillatory boundary-layers \cite{von1974linear,hall1978linear,blennerhassett2002linear}. While such boundary layer flows have been studied extensively in bulk fluids and appear in a wide variety of settings \cite{wang1965flow,riley1966sphere,longuet1953mass,hehner2019stokes,pal2015fluid,ishfaq2019stokes,miller1968oscillations}, their consequences in describing the force on bodies moving along fluid interfaces remain much less well understood. Here, we experimentally examine the unsteady drag and added mass on floating planar bodies undergoing lateral oscillations at an air--water interface, focusing on the regime in which both contributions may be described by an oscillatory boundary layer beneath the body. As a first step, we adopt the simplest theoretical description of a flat plate atop an undeformed interface.

To this end, we employ a magnetic base-excitation platform which drives harmonic oscillations of a floating body translating tangential to an air--water interface. The resulting response amplitude and phase lag, relative to the driving, provide direct access to the effective damping and added mass experienced by the body. This experimental geometry is inspired by classic interfacial rheometry, in which oscillating needles or disks in the creeping-flow limit are used to infer interfacial material properties, including surface viscosity, from their dynamical response \cite{plateau1873statique,marangoni1972principle,manikantan2020surfactant,fitzgibbon2014scaling,braunreuther2023nondestructive,shih2024viscoelastic,zell2016linear}. Here, we use a related configuration for a different purpose: to quantify skin-friction on floating interfacial bodies when fluid inertia is also significant. These measurements allow us to identify the regime in which the simple boundary-layer theory remains predictive for finite-sized bodies, and to quantify its breakdown as additional effects due to interfacial deformation become relevant.

\begin{figure}[t]
	\centering \includegraphics[width=\columnwidth]{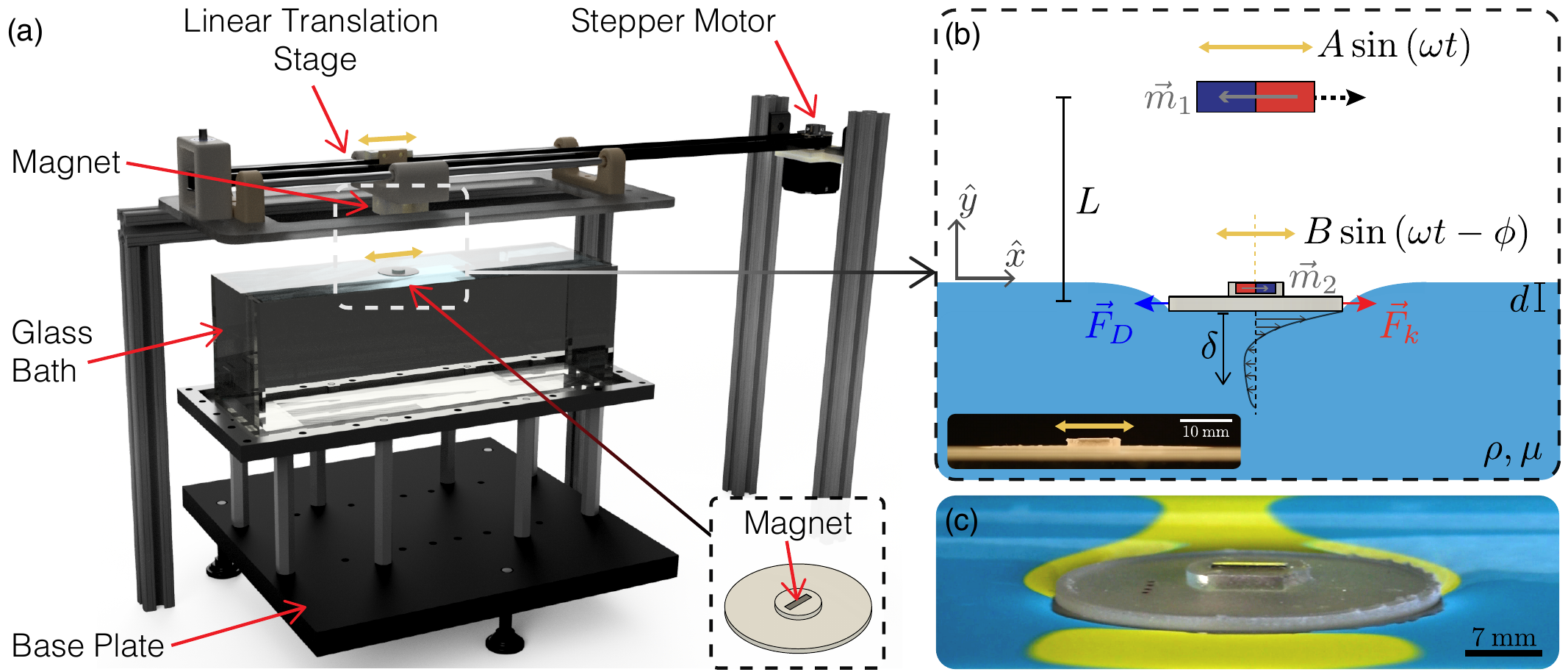}
    \caption{Magnetic-drive platform for measuring unsteady hydrodynamic forces at an air--water interface.
    (a) CAD rendering of the experimental setup. A superhydrophobic disk floats at the surface of a water bath, while a stepper motor drives harmonic oscillations of a linear translation stage carrying a magnet. 
    (b) The larger magnet ($\vec{m}_1$), mounted on the oscillating carriage, has a horizontally oriented dipole moment that is opposite to that of the smaller magnet ($\vec{m}_2$) embedded in a slider. With prescribed displacement $A\sin(\omega t)$ (yellow arrows), the carriage drives the slider, which responds at the same frequency ($\omega=2\pi f$) with amplitude $B$ and phase lag $\phi$. As the carriage moves from left to right (black dotted arrow), the disk is pulled by an effective spring force ($\vec{F}_k$) in the $\hat{x}$ direction, while a boundary layer of characteristic thickness $\delta=\sqrt{\nu/\omega}$ generates a hydrodynamic drag force ($\vec{F}_D$) opposing the motion. Inset: side-view image of the slider on the fluid bath.
    (c) Photograph of the slider floating on water. A reflected yellow and blue background is distorted by the meniscus to produce colors visible at the interface \cite{harris2017visualization}.}
	\label{fig1}
\end{figure}
\section{Experimental Methods}
The experimental setup is shown in Fig.~\ref{fig1}a. A stepper motor drove harmonic oscillations of a linear translation stage carrying a permanent magnet ($|\vec{m}_1| \approx 4.27 \pm 0.08~\mathrm{A\cdot m^2}$) above a glass bath (measuring $29\,\mathrm{cm}$ in length, $9.4\,\mathrm{cm}$ in width, and $9.7\,\mathrm{cm}$ in depth) filled with deionized water at room temperature (density $\rho = 997.4~\mathrm{kg/m^3}$, dynamic viscosity $\mu = 1.002 \times 10^{-3}~\mathrm{Pa \cdot s}$). The bath depth exceeded the oscillatory boundary-layer thickness $\delta$ (Fig.~\ref{fig1}b) by more than two orders of magnitude across the explored frequency range, so bottom-wall confinement effects are negligible. To minimize spurious magnetic interactions, the experimental setup was constructed from non-ferromagnetic materials wherever possible (primarily nylon, aluminum, and SLA-printed parts in Formlabs Clear Resin), the stepper motor was offset $\sim40\,\mathrm{cm}$ from the center of the carriage, and the water surface was raised $\sim30\,\mathrm{cm}$ above the optical table. 

In our experiments, centimeter-scale bodies, hereafter referred to as ``sliders'', of varying shape, size, and mass were manufactured by stereolithographic (SLA) 3D printing (Fig.~\ref{fig1}b,c). The sliders were coated with a commercially available superhydrophobic spray (NeverWet), allowing them to float stably at an air--water interface with the contact line pinned along the lower perimeter $P$, enclosing a planar fluid contact area $a$. In this configuration, the slider of mass $m$ depresses the interface by an amount $d$ under gravity $g$ and is supported by the combined action of hydrostatic pressure and surface tension (Fig.~\ref{fig1}b). A second magnet ($|\vec{m}_2| \approx 0.017 \pm 0.0003~\mathrm{A\cdot m^2}$) embedded in the slider (Fig.~\ref{fig1}a, inset) interacts magnetically with the drive magnet, producing both vertical and horizontal forces. The vertical attractive magnetic force was estimated using a point-dipole approximation, expanding the dipole--dipole interaction about the centered configuration~\cite{griffiths2023introduction}, and was found to be 4--8\% of the slider weight (Supplementary Material~\S II.D). Since this vertical magnetic force is small compared with gravity, $d$ is well approximated by the static vertical force balance $mg \approx \rho g a\, d + \sigma P \sin\theta$, which neglects the upward magnetic attraction. Using the small-slope approximation $\theta \approx d/\ell_c$ gives~\cite{ho2019direct}  
\begin{equation}
    d \approx \frac{mg}{\sigma\left(a/\ell_c^2 + P/\ell_c\right)}
\end{equation}
where $\sigma\approx 0.072\,\mathrm{N/m}$ is the surface tension of an air--water interface, and the capillary length $\ell_c$ is set by a balance between surface tension and gravity,  $\ell_c=\sqrt{\sigma/(\rho g)}$. 

Horizontal forces on the slider arose from the imposed carriage displacement $A\sin(\omega t)$, which drove horizontal oscillations of the slider through an effective lateral spring force $\vec{F}_k$, opposed by a hydrodynamic drag force $\vec{F}_D=F_D\hat{x}$ (Fig.~\ref{fig1}b). The effective horizontal spring constant associated with the magnet-magnet interaction was measured independently in a static experiment using Helmholtz coils to apply a known horizontal force to the floating slider \cite{ho2019direct}. The resulting force--displacement relation was fit to a cubic polynomial and the linear coefficient $k$ was used in the theoretical predictions (Supplementary Material~\S II.A). 

Before each experiment, the bath was cleaned with ethanol and rinsed with deionized water, after which the water level was adjusted to set the vertical separation between the two magnets, fixed at $L = 5.85 \pm 0.05\,\mathrm{cm}$ for all experiments (Fig.~\ref{fig1}b). The magnetic drive and slider response were recorded in side view with a camera aligned with the tank rim (Fig.~\ref{fig1}b, inset) and at $60$\,fps with an image resolution of $\sim15$\,px/mm. Custom image-processing routines written in MATLAB were used to track both the carriage and the slider in each frame (SM, Movie~S1). The driving amplitude, $A$, was obtained from the tracked carriage motion rather than the prescribed stepper kinematics, ensuring that the drive and response measurements are synchronized. The steady-state slider response amplitude, $B$, was determined after the initial transient had decayed, and the phase lag, $\phi$, was obtained from least-squares sinusoidal fits to the steady response, $x_s(t)=B\sin(\omega t-\phi)$ (Fig.~\ref{fig1}b), with discrete Fourier transforms used as an independent consistency check. \markblue{In all experiments, the peak slider speed, $B\omega$, ranged from approximately $0.25$ to $11\,\mathrm{cm\,s^{-1}}$, well below the minimum capillary--gravity wave phase speed at an air--water interface, $c_{\min}\approx23\,\mathrm{cm\,s^{-1}}$ \cite{raphael1996capillary,le2011wave}. Nevertheless, unsteady motion can generate wave drag below this threshold~\cite{closa2010capillary,chepelianskii2008capillary}. Using a linear wave model with weak viscous damping, and modeling the body as a moving pressure distribution as in prior related works, we estimate that the peak wave-induced drag force is only a small fraction (approximately $0.2\%$--$10\%$) of the peak viscous shear force from the oscillatory Stokes layer for the range of parameters considered in our study (Supplementary Material \S V)}. Each reported data point hereafter corresponds to three independent trials performed using three sliders of nearly identical geometry and mass (Fig.~\ref{fig1}c); unless otherwise noted, error bars denote the minimum and maximum across these \markblue{three} trials.

Having defined the experimentally measured response in terms of $A$, $B$, and $\phi$, we now develop a reduced theoretical description for the hydrodynamic force in the limit of a thin oscillatory boundary layer, small amplitude oscillation, and small interfacial deformation.


\section{Oscillatory boundary-layer theory}

To interpret the measured response amplitude and phase lag, we model the slider as a base-excited oscillator whose hydrodynamic forcing arises from the oscillatory viscous boundary layer beneath it. In the section below, we first identify the relevant dimensionless groups, then reduce the Navier--Stokes equations in the thin-layer, small-amplitude limit, and finally derive the corresponding equation of motion and steady harmonic response of the slider.

\subsection{Dimensional analysis}

For all sliders, we define the characteristic length scale as $R = \sqrt{a/\pi}$, which coincides with the disk radius for circular sliders. The sliders oscillate with response amplitude $B$ and angular frequency $\omega$ on a fluid with density $\rho$, dynamic viscosity $\mu$, and kinematic viscosity $\nu=\mu/\rho$. Neglecting explicit dependence on interfacial deformation, dimensional analysis suggests that the added-mass ($m_a$) and damping ($c$) coefficients may be written as
\begin{equation}
    C_A \equiv \frac{m_a}{\rho a R},
    \qquad
    C_D \equiv \frac{c}{\rho a R \omega}, \label{drag_nonD}
\end{equation}
and are a function of two independent dimensionless groups
\begin{equation}
\varepsilon \equiv \frac{B}{R}, \qquad \alpha^{2} \equiv \frac{\omega R^{2}}{\nu}.
\end{equation}
Here $\varepsilon$ is the dimensionless oscillation amplitude, the ratio of the response amplitude $B$ to the body size $R$. Because the slider oscillates in place with no externally imposed flow, the only convective velocity scale is the oscillation velocity itself, $U \sim B\omega$; the importance of advection relative to local unsteadiness is therefore set by $\varepsilon$, \markblue{the characteristic distance a fluid element is displaced during one cycle, measured in body lengths}. Up to a factor of $2\pi$, this is the Keulegan--Carpenter number $KC = U_{\max} T / R = 2\pi B/R$, which conventionally parameterizes oscillatory flows in the absence of a mean current~\cite{sarpkaya2010}. The second group is the Womersley number $\alpha$~\cite{womersley1955method}, with $\alpha = R/\delta$ and $\delta = \sqrt{\nu/\omega}$ the oscillatory viscous penetration depth, so $\alpha$ measures the slider size relative to the boundary-layer thickness and large $\alpha$ corresponds to a thin layer (equivalently, $\alpha^{2}$ is proportional to the frequency parameter $\beta = Re/KC$ of oscillatory-flow theory~\cite{sarpkaya2010}). A Reynolds number may be formed from these, \begin{equation} Re \equiv \frac{B\omega R}{\nu} = \varepsilon\,\alpha^{2}, \end{equation} representing the ratio of fluid inertia to viscous stress. Accordingly, the hydrodynamic response is expected to satisfy $C_A = f_A(\varepsilon,\alpha)$, $C_D = f_D(\varepsilon,\alpha)$. Again, our analysis neglects any explicit role of interfacial deformation, an assumption we show to eventually break down in our experimental measurements.

\subsection{Unsteady limit of the Navier--Stokes equations}
The hydrodynamic response is most tractable in the thin-layer, small-amplitude limit $\alpha \gg 1$, $\varepsilon \ll 1$, which is also the regime accessed by our experiments. When $\varepsilon \ll 1$ the response amplitude is small compared with the body \markblue{size $R$}, so a fluid element is advected only a small fraction of a body length each cycle and advection of momentum across the slider is weak. When $\alpha \gg 1$ viscous momentum diffuses only over a layer thin compared with the body, $\delta/R = \alpha^{-1}\ll1$. We therefore model the near-field flow as a thin oscillatory boundary layer beneath a locally flat interface. For an incompressible fluid, the flow satisfies
\begin{align}
    \nabla \cdot \mathbf{u} &= 0, \\
    \frac{\partial \mathbf{u}}{\partial t} + (\mathbf{u}\cdot\nabla)\mathbf{u}
    &= -\frac{1}{\rho}\nabla p + \nu \nabla^2 \mathbf{u},
\end{align}
where $\mathbf{u}=(u,v)$ are the streamwise and transverse velocity components in the $(x,y)$ plane (Fig.~\ref{fig1}b). We introduce the scalings
\begin{equation}
x=Rx^{\ast},\quad y=\delta y^{\ast},\quad t=\omega^{-1}t^{\ast},\quad
u=Uu^{\ast},\quad v=\tfrac{\delta}{R}Uv^{\ast},\quad p=P\,p^{\ast},
\end{equation}
with streamwise velocity scale $U = B\omega$, and the transverse velocity scale $\frac{\delta}{R}U$ following from continuity. {\color{black} Since the slider oscillates in place with no mean flow, the relevant pressure scale in the boundary layer is set by the viscous shear stresses generated beneath the slider, rather than by a convective dynamic pressure. We therefore take $P \sim \mu U/\delta=\rho \nu B \omega/\delta$.} 
Substituting {\color{black} these scalings} into the transverse ($y$) momentum equation and dividing through by the
pressure-gradient scale gives
{\color{black}
\begin{equation}
\frac{1}{\alpha}\frac{\partial v^{*}}{\partial t^{*}}
+\frac{\varepsilon}{\alpha}\!\left(u^{*}\frac{\partial v^{*}}{\partial x^{*}}
+v^{*}\frac{\partial v^{*}}{\partial y^{*}}\right)
=-\frac{\partial p^{*}}{\partial y^{*}}
+\frac{1}{\alpha^{3}}\frac{\partial^{2} v^{*}}{\partial x^{*2}}
+\frac{1}{\alpha}\frac{\partial^{2} v^{*}}{\partial y^{*2}}.
\end{equation}}
{\color{black}So for $\epsilon\ll1$ and $\alpha\gg1$, the leading-order expression is,}
\begin{equation}
\frac{\partial p^{\ast}}{\partial y^{\ast}} = 0.
\end{equation}

{\color{black} The pressure is therefore uniform across the boundary layer. Similarly, substituting the characteristic scales mentioned above into the streamwise ($x$) momentum equation and dividing by $U \omega$ gives, 
\begin{equation}
\frac{\partial u^{\ast}}{\partial t^{\ast}}
+\varepsilon\!\left(u^{\ast}\frac{\partial u^{\ast}}{\partial x^{\ast}}
+v^{\ast}\frac{\partial u^{\ast}}{\partial y^{\ast}}\right)
=-\frac{1}{\alpha}\frac{\partial p^{\ast}}{\partial x^{\ast}}
+\frac{1}{\alpha^{2}}\frac{\partial^{2} u^{\ast}}{\partial x^{\ast 2}}
+\frac{\partial^{2} u^{\ast}}{\partial y^{\ast 2}}.
\end{equation}

}

{\color{black}Taking the same limit $\varepsilon \ll 1$ and $\alpha \gg 1$, and assuming that no imposed streamwise pressure gradient of size $O(\alpha)$ is present, the streamwise advection, streamwise pressure-gradient, and streamwise viscous-diffusion terms are subleading. The leading-order balance is therefore between local unsteadiness and wall-normal viscous diffusion. Recasting into dimensional
variables recovers the diffusion equation,}
\begin{equation}\label{diffusion}
\frac{\partial u}{\partial t}=\nu\frac{\partial^{2} u}{\partial y^{2}}.
\end{equation}

\subsection{Equation of motion of the slider}
We now couple the reduced fluid model to the solid-body dynamics of the floating slider. Let $x_s(t)$ denote the slider displacement along the $x$-axis, tangent to the fluid interface. Linearizing the magnetic interaction about equilibrium gives an effective spring of stiffness $k$, which enters through the spring force $-k(x_s-x_d)$, and the slider may be modeled as a base-excited oscillator,
\begin{equation}
    m\ddot{x}_s + kx_s = kA\sin(\omega t) + F_D(t),
\end{equation}
where $x_d=A\sin(\omega t)$ is the prescribed motion of the driving magnet, and $F_D(t)$ is the signed hydrodynamic force exerted by the fluid on the slider. We will show below that at steady state, $F_D(t)$ admits a decomposition into its reactive ($m_a \ddot{x}_s$) and dissipative ($c\dot{x}_s$) components, yielding the compact form
\begin{equation}\label{eqn_of_motion}
(m+m_a)\ddot{x}_s+kx_s+c\dot{x}_s=kA\sin(\omega t).
\end{equation}

To obtain the coefficients $m_a$ and $c$, we first compute the instantaneous $F_D(t)$ by solving equation~\eqref{diffusion} in the semi-infinite domain $y\le 0$, subject to the boundary conditions
\begin{equation}
    u(0,t)=\dot{x}_s(t), 
    \qquad
    u(y\to-\infty,t)=0,
\end{equation}
together with the initial condition
\begin{equation}
    u(y,0)=0
\end{equation}
where we have assumed that the slider is an infinite plate, and that the fluid is initially quiescent and driven entirely by the time-dependent plate motion. This is the classical unsteady Stokes problem \cite{stokes1922mathematical}, and the wall shear stress for this unsteady viscous diffusion problem can be written as a history integral \cite{schlichting2016boundary}, \begin{equation}
    \tau(t)
    =
    \mu \left.\frac{\partial u}{\partial y}\right|_{y=0}
    =
    \sqrt{\frac{\rho\mu}{\pi}}
    \int_0^t
    \frac{\ddot{x}_s(t')}{\sqrt{t-t'}}\,dt'.
\end{equation}
For a prescribed sinusoidal wall velocity, the history form yields a wall shear in the fluid that agrees with the exact Laplace-transform solution, including the start-up transient~\cite{erdogan2000note}.
Approximating the total force on the slider as the shear stress integrated over the slider's fluid contact area $a$ gives
\begin{equation}
    F_D(t)
    =
    -a\sqrt{\frac{\rho\mu}{\pi}}
    \int_0^t
    \frac{\ddot{x}_s(t')}{\sqrt{t-t'}}\,dt'.
\end{equation}
The full unsteady equation of motion is therefore
\begin{equation}\label{basset}
    m\ddot{x}_s
    + kx_s
    + a\sqrt{\frac{\rho\mu}{\pi}}
    \int_0^t
    \frac{\ddot{x}_s(t')}{\sqrt{t-t'}}\,dt'
    =
    kA\sin(\omega t).
\end{equation}

This form retains the transient memory of the viscous boundary layer during start-up. 

\subsection{Steady-state response}
At long times ($t \rightarrow \infty$) equation \eqref{basset} reduces to the familiar oscillatory Stokes drag \cite{schlichting2016boundary},
\begin{equation}
    F_{D,\infty}(t)
    =
    -a\sqrt{\rho\mu\omega}\,B\omega
    \cos\!\left(\omega t-\phi+\frac{\pi}{4}\right),
\end{equation}
for the harmonic steady-state response $x_s(t)=B\sin(\omega t-\phi)$. Using the identity $\cos\!\left(\theta+\frac{\pi}{4}\right)=
    (\cos\theta-\sin\theta)/\sqrt{2}$, the drag can be expressed
\begin{equation}
    F_{D,\infty}(t)=-c\dot{x}_s-m_a\ddot{x}_s,
\end{equation}
with
\begin{equation}
    c=\frac{a\sqrt{\rho\mu\omega}}{\sqrt{2}},
    \qquad
    m_a=\frac{a\sqrt{\rho\mu\omega}}{\sqrt{2}\,\omega}
    =
    \frac{\rho a \delta}{\sqrt{2}}.\label{eqn:cma}
\end{equation}
representing the steady-state damping and added mass respectively. \markblue{The physical origin of these expressions can also be understood directly from the characteristic oscillatory boundary-layer thickness. The volume of fluid accelerated beneath the slider scales as $a\delta$, so the associated added mass scales as $m_a\sim\rho a\delta$, in agreement with the exact result above. Likewise, for a characteristic slider speed $U$, the velocity gradient across the layer scales as $U/\delta$, giving a shear stress $\tau\sim\mu U/\delta$. Integrating this stress over the contact area $a$ and using $\delta=\sqrt{\nu/\omega}$ gives $F_\mu\sim a \sqrt{\rho \mu \omega}U$ such that $c\sim a\sqrt{\rho\mu\omega}$, similarly in agreement with the corresponding exact result. 

Substituting the results from equation \eqref{eqn:cma} into equation \eqref{drag_nonD} gives the final results for the added-mass and damping coefficients,
    \begin{equation}
    \begin{aligned}
    C_A \equiv \frac{m_a}{\rho a R}
    &=
    \frac{\rho a\delta/\sqrt{2}}{\rho aR}
    =
    \frac{\delta}{\sqrt{2}R}
    =
    \frac{1}{\sqrt{2}\alpha},
    \\
    C_D \equiv \frac{c}{\rho aR\omega}
    &=
    \frac{a\sqrt{\rho\mu\omega}/\sqrt{2}}{\rho aR\omega}
    =
    \frac{\delta}{\sqrt{2}R}
    =
    \frac{1}{\sqrt{2}\alpha}.
    \end{aligned}
    \label{coefficient}
    \end{equation}}
 \markblue{Both coefficients therefore scale linearly with the inverse Womersley number, as we later compare with the experimentally inferred values}. The corresponding steady-state response of the slider follows from solving equation \eqref{eqn_of_motion}. Defining the natural frequency $\omega_n \equiv \sqrt{k/m}$, $r \equiv \omega/\omega_n$,
the long-time solution is
\begin{equation}\label{sol1}
    \frac{B}{A}
    =
    \frac{1}{r^2\sqrt{c_1^2+c_2^2}},
    \qquad
     \markblue{\phi =
    \arccos\left(
    \frac{c_1}{\sqrt{c_1^2+c_2^2}}
    \right)}.
\end{equation}
where
\begin{equation}\label{sol2}
    c_1 \equiv -1 + \frac{1}{r^2} - \frac{m_a}{m},
    \qquad
    c_2 \equiv \frac{c}{m\omega}.
\end{equation}

\section{Results}
\begin{figure}[t]
	\centering
    \includegraphics[width=\columnwidth]{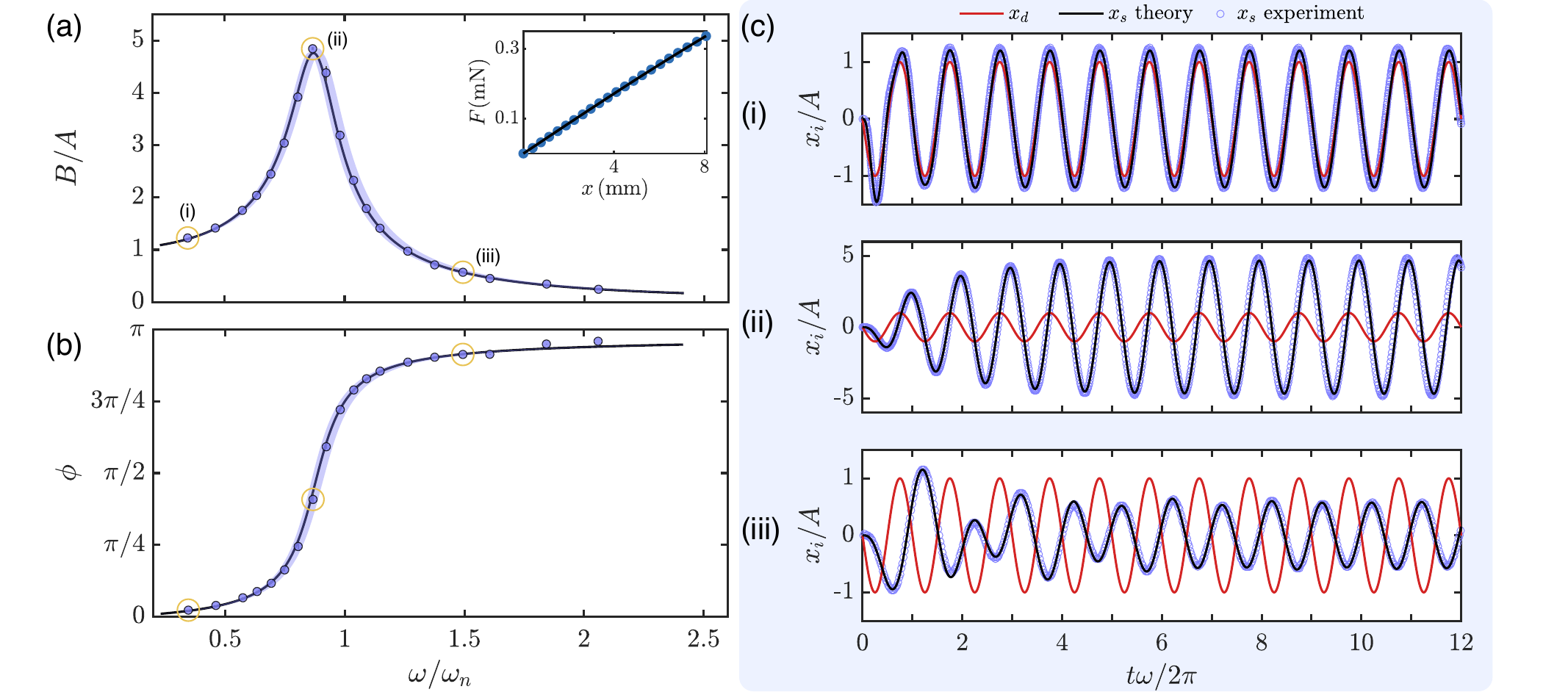}
	\caption{Frequency response of a magnetically driven slider at an air--water interface.
    (a) Normalized response amplitude, $B/A$, of a circular slider of radius $R = 20 \pm 0.1\,\mathrm{mm}$ and mass $m = 1.49 \pm 0.02\,\mathrm{g}$ driven over a range of frequencies. Circles denote experiments; error bars show the minimum and maximum over three trials using three disks of identical mass and geometry. The solid curve is the theoretical prediction obtained by assuming an oscillatory boundary layer (equation \ref{sol1}-\ref{sol2}), and the shaded region indicates uncertainty arising from small variations in slider mass and the measured spring constant. Inset: the effective spring constant, $k$, is measured by horizontally displacing the slider with Helmholtz coils that apply a known horizontal force (Supplementary Material~\S II.A).
    (b) Corresponding phase lag, $\phi$, as a function of frequency. 
    (c) Time-series response of the slider $x_s(t)$ (circles) compared with the prediction from equation \eqref{basset} (black line), which additionally predicts the early time transients. The imposed motion of the drive $x_d(t)$ (red line) is overlaid to highlight the phase lag. (i) $f=0.3\,\mathrm{Hz}$; (ii) $f=0.75\,\mathrm{Hz}$; and (iii) $f=1.3\,\mathrm{Hz}$. All experiments were performed at high Womersley number and small amplitude ratio: $\alpha=27$--$67$ and $\varepsilon = B/R = 0.05$--$0.40$, with $Re=\varepsilon\alpha^{2}=210$--$550$. The driving amplitude was varied over the range $A=1.2$--$4.8$~mm.}
	\label{fig2}
\end{figure}
We begin by comparing the predictions of the idealized oscillatory boundary-layer model (equations~\ref{sol1}, \ref{sol2}) against the measured frequency response of a circular slider. Figure~\ref{fig2}(a,b) shows the normalized response amplitude, $B/A$, and phase lag, $\phi$, for a slider of radius $R=20\,\mathrm{mm}$ and mass $m=1.49\,\mathrm{g}$ driven over a range of frequencies. When the driving frequency is normalized by the natural frequency, the data exhibit the expected resonance peak in $B/A$ near the natural frequency, together with a transition in the phase shift from in-phase oscillations ($\phi \approx 0$) to antiphase oscillations ($\phi \approx \pi$) as $\omega$ increases. \markblue{In the low-frequency limit, the spring dominates the response, and the slider follows the drive quasi-statically, such that $B/A \rightarrow 1$ and $\phi \rightarrow 0$. In the high-frequency limit, the inertial contribution $-(m+m_a)\omega^2$ dominates the elastic and viscous contributions $k$ and $c\omega$, respectively, such that $B/A\rightarrow 0$ and $\phi\rightarrow\pi$ (SM, Movie~S2).} The theoretical prediction obtained from the simple model is in good agreement with the experimental data, with no fit parameters; the effective spring constant, $k$, is measured independently for each slider (Fig.~\ref{fig2}(a), inset, Supplementary Material~\S II.A). Across all frequency-sweep
experiments, the relative displacement $q = x_s - x_d$ between the slider
and the carriage remains small (the average of the maximum $|q|$ measured in each trial was
$\langle \max_t |q(t)| \rangle_{\mathrm{trials}} \approx 5.4\,\mathrm{mm}$, while the largest value observed
over all trials was $\max_{\mathrm{trials},t}|q(t)| \approx 11.3\,\mathrm{mm}$) and the resulting effect of
the linear-spring approximation is quantified in Supplementary
Material \S II.B,C. The shaded band around the theoretical curve reflects propagated uncertainty in the measured spring constant and slider mass. Across these experiments, the flow remains in the high-$\alpha$, low-$\varepsilon$ regime, consistent with the assumptions underlying the oscillatory Stokes-layer model.

Figure~\ref{fig2}(c) further compares the measured time series with the solution of equation~\eqref{basset}. For this comparison, the integro-differential equation was solved numerically using a fourth-order Adams--Bashforth time-stepping scheme, with the history integral evaluated at each time step by Simpson's rule. Simulations were initialized from rest, using the independently measured values of $A$, $m$, and $k$ together with the fluid properties, and integrated with a fixed time step $\Delta t = 10^{-3}\,\mathrm{s}$. In addition to reproducing the steady-state amplitude and phase lag, the model captures the transient approach to the periodic response after the motion is initiated. This agreement further supports the interpretation that the dominant hydrodynamic resistance in this regime is set by the unsteady viscous boundary layer beneath the slider (see Supplementary Material \S IV for additional time-series comparisons across shapes, sizes, and deformation levels). Remaining differences between experiment and theory may reflect unmodeled contributions from interfacial deformation and finite-geometry effects, both of which are examined in more detail below.

\begin{figure}[t]
	\centering
    \includegraphics[width=\columnwidth]{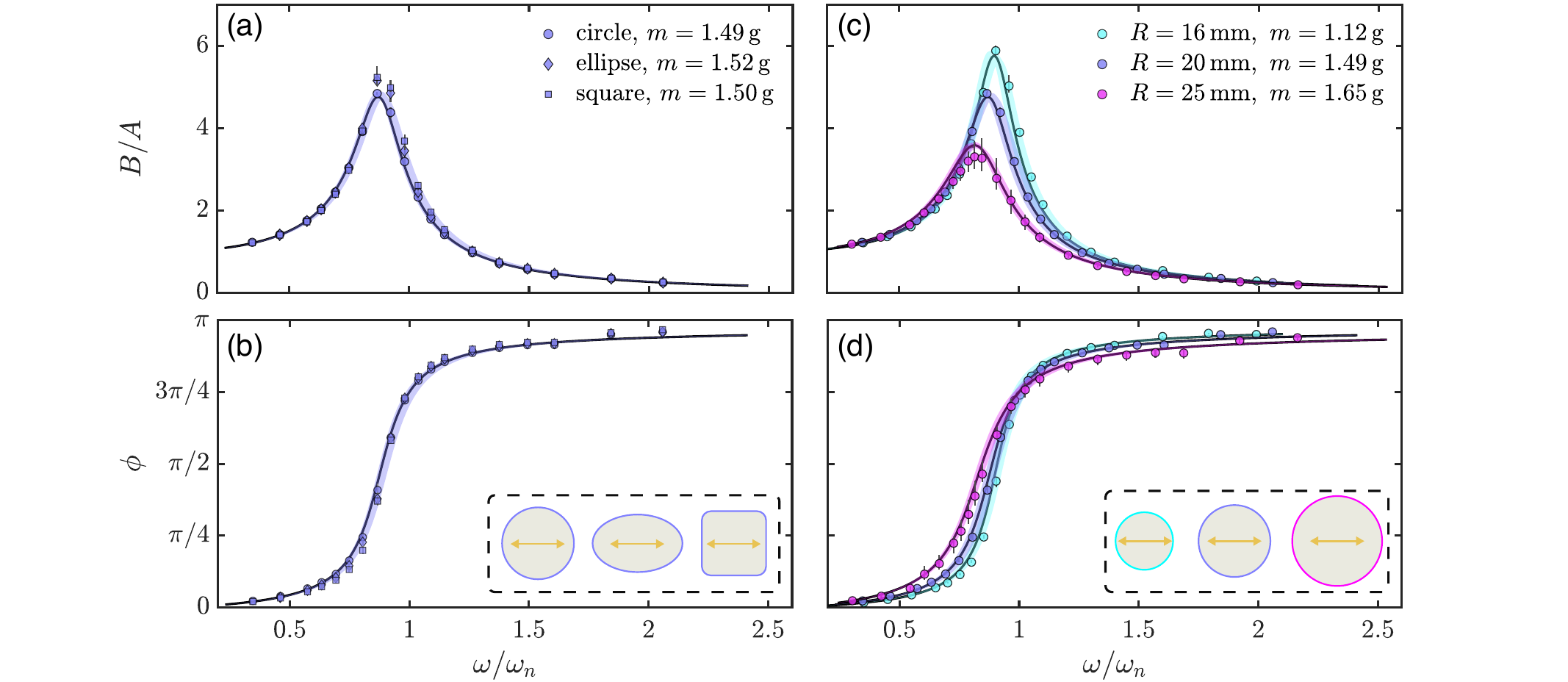}
	\caption{Frequency response for sliders of various shapes, sizes, and masses.
    (a,b) Frequency sweeps of the normalized amplitude response and phase lag for sliders of different shape but equal area $a = 1257 \pm 12\,\mathrm{mm^2}$ (see Supplementary Material~\S VIII, Table~S2 for shape dimensions). Because the theory computes the hydrodynamic forcing by integrating the shear stress over the slider contact area, it predicts the same response curve for all shapes. The shaded region indicates uncertainty due to small variations in the measured stiffness and slider mass. Error bars show the minimum and maximum \markblue{across three independent trials using three sliders}.
    (c,d) Frequency response for circular sliders with varying radius and mass. Across all experiments, $\alpha=24$--$84$ and $\varepsilon = B/R = 0.03$--$0.56$, with $Re=\varepsilon\alpha^{2}=200$--$620$. The driving amplitude was varied over the range $A=1.0$--$7.1$~mm.}
	\label{fig3}
\end{figure}

Having established agreement for a single circular slider, we next test the extent to which the theory generalizes across slider geometry, mass ($m$), and contact area size ($a$). Figure~\ref{fig3}(a,b) compares sliders of different shape but equal contact area and mass. Because the theory computes the net hydrodynamic forcing by integrating the shear stress over the slider contact area, it predicts identical response curves for all equal-area sliders (equation \ref{sol1}--\ref{sol2}). The measured amplitude and phase data collapse closely onto this common prediction, indicating that three-dimensional shape-dependent effects are weak in the present parameter regime. Figure~\ref{fig3}(c,d) then shows frequency sweeps for circular sliders with different radii and masses. As the size and mass of the slider are varied, the resonance location and peak amplitude shift accordingly, but the reduced model continues to capture the observed trends over the full ensemble of experiments.

\begin{figure}[t]
	\centering
    \includegraphics[width=\columnwidth]{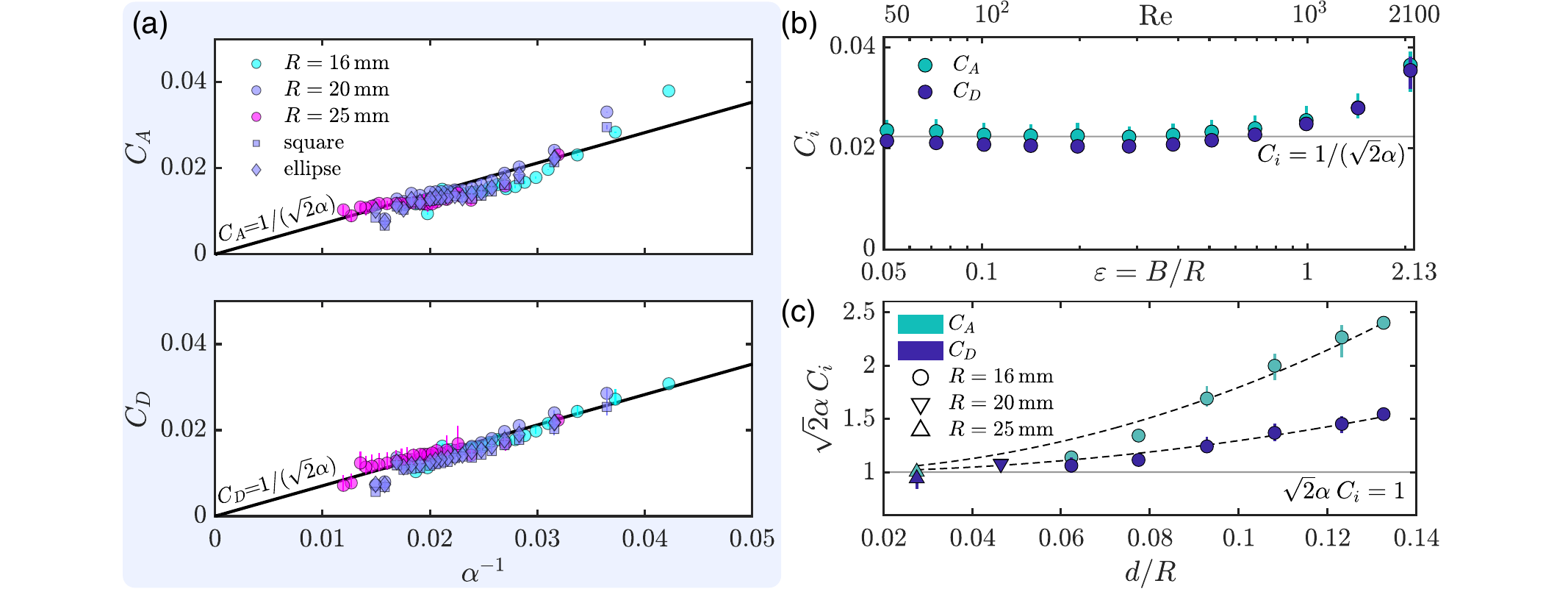}
	\caption{Added-mass and damping coefficients inferred from the measured response amplitude and phase lag.
    (a) Added-mass coefficient (top) and damping coefficient (bottom) plotted against the inverse Womersley number. The combined data from Fig.~\ref{fig3} collapse across sliders of different radius $R$, mass $m$, and shape, with each point denoting the mean over three trials using three sliders of identical geometry and mass. Unsteady boundary-layer theory predicts a linear dependence of both coefficients on $\alpha^{-1}$ and is shown by the solid curves.
    (b) Deviations from the theory emerge at large amplitude ratio $\varepsilon$. A circular slider with $R=20\,\mathrm{mm}$ and $m=1.49\,\mathrm{g}$ is driven at fixed frequency, $f=0.4\,\mathrm{Hz}$, while the driving amplitude is varied, producing a range of amplitude ratios $\varepsilon = B/R$ (hence a range of $Re=\varepsilon\alpha^{2}$ at fixed $\alpha$). Because $\alpha$ is fixed, the theory predicts constant coefficients; here $\alpha=31.6\pm0.2$.
    (c) Deviations from the theory also emerge when interfacial deformation becomes large. The mass and radius of circular sliders are varied, and the predicted quasistatic interfacial deformation, normalized by the disk radius, is plotted. While $\sqrt{2}\alpha C_i=1$ remains constant ($\alpha=31.4\pm0.2$), the experimental points depart from the prediction at large deformations. \markblue{Dashed curves are fits to $\sqrt{2}\alpha C_i=1+a_i(d/R)^2$; at fixed $\alpha$ and $Bo=\rho gR^2/\sigma$, the meniscus scaling predicts $a_A=\sqrt{2}\alpha K_A$ and $a_D=\sqrt{2}BoK_D/\alpha$. Separate fits of the meniscus scalings give $K_A\approx1.80$ and $K_D\approx18.4$}. Error bars show the minimum and maximum over \markblue{three} trials.}
	\label{fig4}
\end{figure}

Beyond validating the frequency-response curves, the steady-state measurements may also be inverted to infer the effective added-mass and damping coefficients directly. Inverting equations \eqref{sol1}-\eqref{sol2} for the experimentally measured $B/A$ and $\phi$ yields the dimensional coefficients $m_a$ and $c$ through $c_1 = A\cos\phi/Br^2$ and $c_2 = A\sin\phi/Br^2$. The corresponding nondimensional coefficients are
\begin{equation}
   C_A
    = \frac{m_a}{\rho a R}
    = \frac{m}{\rho a R}\left(-1+\frac{1}{r^2}-c_1\right),
    \qquad
    C_D
    = \frac{c}{\rho a R \omega}
    = \frac{m\omega c_2}{\rho a R \omega}
    = \frac{m}{\rho a R}\,c_2.\label{inversion}
\end{equation}
Figure~\ref{fig4}(a) shows that when $C_A$ and $C_D$ are plotted against $\alpha^{-1}$, the data collapse across sliders of different radius, mass, and shape onto the linear scaling predicted by the oscillatory boundary-layer theory (equation \ref{coefficient}). Despite the geometric and inertial variations across the ensemble of sliders, the effective hydrodynamic coefficients are nearly collapsed by the inverse Womersley number, as suggested by the theory. Additional frequency-sweep results, including heavier sliders, are presented in Supplementary Material~\S III.

Figures~\ref{fig4}(b,c) explore and identify some of the limits of the simple model. In Fig.~4(b), the driving amplitude is varied at fixed frequency for a circular slider with radius $R=20$~mm, so that the response amplitude $B$ changes while $\alpha^{2}=\omega R^{2}/\nu$ remains fixed. Thus, varying $B$ varies the amplitude ratio $\varepsilon = B/R$, and hence $Re=\varepsilon\alpha^{2}$, at fixed $\alpha$. In this case the theory predicts constant coefficients $(C_A,C_D)$, and the systematic rise in both $C_A$ and $C_D$ as $\varepsilon$ increases indicates the onset of advective effects beyond the oscillatory Stokes-layer approximation. We note that the weak nonlinearity of the magnetic spring produces a small upward shift in $C_A$ at the largest response amplitudes, but this correction is too modest to account for the full deviation observed at large amplitude in experiment (Supplementary Material~\S II.C).

In Fig.~\ref{fig4}(c), we examine deviations associated with interfacial deformation by plotting $\sqrt{2}\alpha C_i$ against the predicted quasistatic deformation, $d$, normalized by the slider radius for circular sliders, where $i\in\{A,D\}$. The geometric factor $d/R$ represents the aspect ratio of the interfacial depression created by the slider in equilibrium. The rescaling of the coefficients allows sliders of different radii to be compared on the same axes, since in the oscillatory Stokes-layer limit $\alpha C_i=1/\sqrt{2}$ is independent of $\alpha$ (equation \ref{coefficient}). To probe this effect directly, we performed additional experiments with a slider of radius $R=16\,\mathrm{mm}$ by incrementally adding washers to increase the mass and thus the static interfacial deformation (mass range $m = 1.10$--$2.33\,\mathrm{g}$, corresponding to $d/R \approx 0.06$--$0.13$). The data depart systematically from the constant-$\sqrt{2}\alpha C_i$ prediction as $d/R$ increases, with $\sqrt{2}\alpha C_A$ and $\sqrt{2}\alpha C_D$ both appearing to increase with deformation. The increase is notably steeper for $\sqrt{2}\alpha C_A$ than for $\sqrt{2}\alpha C_D$, suggesting that interfacial deformation affects the added mass more strongly than the drag. The results from Fig.~\ref{fig3}(c,d) for circular sliders of different radii are also plotted, and lie in the low-interfacial deformation regime. Taken together, these results show that the oscillatory boundary-layer theory accurately captures the hydrodynamic forcing when the interfacial deformation remains small (i.e. $d/R\ll1$), while also clearly delineating the regimes in which additional physics must be incorporated.
\section{Discussion}
The results presented suggest that the dominant horizontal hydrodynamic resistance on the oscillating sliders is viscous skin friction generated within an oscillatory boundary layer beneath the body. In the regime of focus explored herein, the measured frequency-response curves and the inferred hydrodynamic coefficients are both consistent with an idealized oscillatory boundary-layer prediction, over a range of slider radii, masses, and shapes. The present work extends the quasi-steady picture established previously for freely decelerating sliders \cite{pucci2019friction}, whose associated fluid dynamics were modeled as a quasi-static Blasius boundary layer, to the unsteady regime. More generally, the quasi-steady and unsteady descriptions can be viewed as complementary limits of laminar boundary-layer theory: when a weak, rapid streamwise oscillation is combined with a high-Reynolds-number Blasius flow, the flow asymptotically separates into an outer modified Blasius layer that convects the mean flow and an inner Stokes shear layer that accommodates the oscillatory motion. The two regions are then matched asymptotically to obtain the complete flow field and, consequently, the associated drag ~\cite{ackerberg1972unsteady,rott1960response,lighthill1954response}. Such combined translation and oscillation also characterizes self-propelled interfacial bodies~\cite{rhee2022surferbot}, motivating extension of the present framework to that regime. \markblue{For example, estimates based on representative local scales for a miniature ambulatory robot~\cite{chen2018controllable} and honeybee wings~\cite{roh2019honeybees} give $(\alpha,\epsilon)\approx(28,0.20)$ and $(80,0.12)$, respectively, within the ranges explored here ($\alpha=24$--$84$, $\epsilon=0.03$--$0.56$; see Supplementary Material~\S VII, Table~S1).}

The agreement between the simple theory and the measurements is notable, given that the model is based on Stokes' second problem for a flat, unbounded plate and therein neglects finite-size and three-dimensional effects. This agreement may be rationalized in terms of the non-dimensional parameters governing the problem. Firstly, the experiments are conducted in a regime where the Womersley number $\alpha$ is large, indicating that the viscous penetration depth $\delta = \sqrt{\nu/\omega}$ remains small compared to the slider size $R$.  As such, only a thin layer of fluid beneath the body is sheared during each cycle.

In addition, operating at sufficiently small amplitude ratio $\varepsilon \ll 1$ implies that the motion reverses before flow disturbances associated with the disruptive leading and trailing edges can propagate across the slider. Lastly, working with light sliders where the interfacial depression $d$ is small relative to the body size $R$ results in a geometric configuration most akin to a flat planar boundary.  The combination of these conditions (i.e., $\alpha\gg1,\varepsilon \ll 1,$ and $d/R\ll1$) helps explain why the experiments collapse onto a common response that is well predicted by the simplified two-dimensional theory (Fig.~\ref{fig4}a). Furthermore, we operate in a regime where the oscillatory boundary layer is predicted to be stable.  Using the characteristic slider speed $U_0 = B\omega$, the corresponding Stokes-layer Reynolds number is
$
Re_{\delta}\equiv\frac{U_{0}}{\sqrt{2\nu\omega}}
=\frac{\varepsilon\,\alpha}{\sqrt{2}}
\approx 1.7\text{--}13.1,
$
across our main ensemble, well below the flat-Stokes-layer linear-instability threshold of $Re_{\delta,c} \approx 700$ \cite{blennerhassett2002linear}.

The systematic deviations documented in Fig.~\ref{fig4}(b,c) further clarify the limits of this reduced description, but nevertheless demonstrate the utility of the experimental method outside the particular asymptotic regime. At fixed $\alpha$, increasing $\varepsilon$ drives the system away from the oscillatory Stokes-layer limit, suggesting that advective effects associated with the finite plate size begin to measurably influence the stresses on the plate. In particular, both the apparent added mass and damping increase as $\varepsilon$ is increased. Likewise, increasing the interfacial deformation produces systematic departures from the theoretical prediction, with $C_A$ increasing more rapidly than $C_D$. This result suggests that interfacial deformation affects the reactive
component of the hydrodynamic force more strongly than the dissipative
component. The advected meniscus may contribute an additional added-mass
effect not captured by the flat-interface model, while also introducing
extra dissipation through near-edge flow such as contact-line motion or
local recirculation at the pinned rim. We visualize this near-edge flow
qualitatively for a highly deformed case, observing fluid motion along the
deformed meniscus (SM Movie~S3).

\markblue{To interpret these trends, we estimate the additional added mass from the acceleration-induced asymmetry of the meniscus in Supplementary Material \S VI. Using the unsteady Bernoulli relation in the interfacial normal pressure balance gives a net horizontal capillary force opposing the slider acceleration. We estimate the additional damping from viscous dissipation in the outer potential flow near the corner formed by the slider underside and the meniscus. Expanding about a flat interface suggests leading-order meniscus corrections to the oscillatory Stokes-layer prediction, $C_A \simeq 1/(\sqrt{2}\alpha)+K_A(d/R)^2$ and $C_D \simeq 1/(\sqrt{2}\alpha) +K_D\mathrm{Bo}\,\alpha^{-2}(d/R)^2$. Here $\mathrm{Bo}=\rho gR^2/\sigma=(R/\ell_c)^2$ is the Bond number, measuring the relative importance of gravity and surface tension. The coefficients $K_A$ and $K_D$ are dimensionless prefactors that cannot be determined from scaling alone. These scaling estimates provide a qualitative interpretation of the onset of departures from the flat-interface prediction as $d/R$ increases.}

More broadly, our work establishes a simple and accessible non-contact method to quantify unsteady hydrodynamic forces at fluid interfaces. Because added mass is an acceleration-proportional contribution whereas viscous drag is velocity-proportional, steady or quasi-steady drag measurements at fluid interfaces \cite{pucci2019friction,petkov1995measurement,hunt2023drag,kamoliddinov2021hydrodynamic} typically do not facilitate the characterization of both the reactive and dissipative parts of the loading inherent to an unsteady problem. The frequency-response framework is thus an alternative to steady drag measurements: resolving both amplitude and phase yields the effective added-mass and damping coefficients simultaneously. Our experimental framework is also well suited for studying fluid-solid interaction and unsteady hydrodynamics on the interface of complex media, including non-Newtonian \cite{ai2005investigation,fetecau2009note}, active \cite{foffano2012colloids,knevzevic2021oscillatory}, biological \cite{ruhs2013situ}, or odd-viscous \cite{soni2019odd} fluid interfaces where interfacial microstructure and additional physics can qualitatively reshape unsteady drag and added mass. In this way, the driven motion of a floating body may offer a direct probe of the unsteady mechanics of fluid interfaces.

\begin{acknowledgments}
D.M.H. gratefully acknowledges financial support from the National Science Foundation (CBET-2338320) and the Office of Naval Research (N00014-21-1-2816 and N00014-21-1-2670). The authors would like to thank Luke Alventosa and Giuseppe Pucci for fruitful discussions. I.H. also acknowledges Manu Prakash for helpful feedback and support during manuscript preparation. A.H.K. acknowledges Osman A. Basaran for valuable suggestions and support during the preparation of the manuscript.
\end{acknowledgments}

\section*{Author Declarations}
The authors report no conflicts of interest.

\section{Data Availability}
The open-source data release contains the full experimental dataset for oscillating interfacial sliders across multiple geometries and forcing conditions, including circular sliders of radius $R=16$, $20$, and $25\,\mathrm{mm}$, square and elliptical sliders, and additional $R=16\,\mathrm{mm}$ cases with added weight. The dataset is organized into seven MATLAB \texttt{.mat} files, each containing a structure array in which each element corresponds to one experimental trial. For each trial, the stored variables include the driving frequency $f$, the driving amplitude $A$, the measured slider amplitude $B$, the response ratio $B/A$, the phase lag $\phi$ (reported in both radians and degrees), a trial index, and the full time-resolved records of time and slider position, $t$ and $x_{\mathrm{slider}}(t)$. The released data files and example plotting script are publicly available at \url{https://github.com/harrislab-brown/Oscillating-Sliders}.

\bibliography{references}

@article{jung2026ground,
  title={Ground effect on Undulation and pumping near surfaces},
  author={Jung, Sunghwan},
  journal={Integrative and Comparative Biology},
  volume={66},
  pages={icag026},
  year={2026},
  publisher={Oxford University Press}
}

@article{tarr2024probing,
  author  = {Tarr, S. W. and Brunner, J. S. and Soto, D. and Goldman, D. I.},
  title   = {Probing hydrodynamic fluctuation-induced forces with an oscillating robot},
  journal = {Phys. Rev. Lett.},
  volume  = {132},
  number  = {8},
  pages   = {084001},
  year    = {2024},
}

@article{o2026achieving,
  author  = {O'Donovan, D. and Bustamante, M. D. and Devauchelle, O. and Benham, G. P.},
  title   = {Achieving optimal locomotion using self-generated waves},
  journal = {J. Fluid Mech.},
  volume  = {1029},
  pages   = {A4},
  year    = {2026},
}

@article{sungar2025synchronization,
  author  = {Sungar, N. and Sharpe, J. and Ijzerman, L. and Barotta, J.-W.},
  title   = {Synchronization and self-assembly of free capillary spinners},
  journal = {Phys. Rev. E},
  volume  = {111},
  number  = {3},
  pages   = {035104},
  year    = {2025},
}

@article{hartmann2025highly,
  author  = {Hartmann, F. and Baskaran, M. and Raynaud, G. and Benbedda, M. and Mulleners, K. and Shea, H.},
  title   = {Highly agile flat swimming robot},
  journal = {Sci. Robot.},
  volume  = {10},
  number  = {99},
  pages   = {eadr0721},
  year    = {2025},
}

@article{roh2019honeybees,
  author  = {Roh, C. and Gharib, M.},
  title   = {Honeybees use their wings for water surface locomotion},
  journal = {Proc. Natl. Acad. Sci. U.S.A.},
  volume  = {116},
  number  = {49},
  pages   = {24446--24451},
  year    = {2019},
}

@article{benham2024wave,
  author  = {Benham, G. P. and Devauchelle, O. and Thomson, S. J.},
  title   = {On wave-driven propulsion},
  journal = {J. Fluid Mech.},
  volume  = {987},
  pages   = {A44},
  year    = {2024},
}

@article{harris2017visualization,
  author  = {Harris, D. M. and Quintela, J. and Prost, V. and Brun, P.-T. and Bush, J. W. M.},
  title   = {Visualization of hydrodynamic pilot-wave phenomena},
  journal = {J. Vis.},
  volume  = {20},
  number  = {1},
  pages   = {13--15},
  year    = {2017},
}

@article{pucci2019friction,
  author  = {Pucci, G. and Ho, I. and Harris, D. M.},
  title   = {Friction on water sliders},
  journal = {Sci. Rep.},
  volume  = {9},
  number  = {1},
  pages   = {4095},
  year    = {2019},
}

@article{ho2019direct,
  author  = {Ho, I. and Pucci, G. and Harris, D. M.},
  title   = {Direct measurement of capillary attraction between floating disks},
  journal = {Phys. Rev. Lett.},
  volume  = {123},
  number  = {25},
  pages   = {254502},
  year    = {2019},
}

@article{ho2023capillary,
  author  = {Ho, I. and Pucci, G. and Oza, A. U. and Harris, D. M.},
  title   = {Capillary surfers: Wave-driven particles at a vibrating fluid interface},
  journal = {Phys. Rev. Fluids},
  volume  = {8},
  number  = {11},
  pages   = {L112001},
  year    = {2023},
}

@article{oza2023theoretical,
  author  = {Oza, A. U. and Pucci, G. and Ho, I. and Harris, D. M.},
  title   = {Theoretical modeling of capillary surfer interactions on a vibrating fluid bath},
  journal = {Phys. Rev. Fluids},
  volume  = {8},
  number  = {11},
  pages   = {114001},
  year    = {2023},
}

@article{harris2025propulsion,
  author  = {Harris, D. M. and Barotta, J.-W.},
  title   = {Propulsion and interaction of wave-propelled interfacial particles},
  journal = {Phys. Rev. Fluids},
  volume  = {10},
  number  = {10},
  pages   = {100503},
  year    = {2025},
}

@article{barotta2025synchronization,
  author  = {Barotta, J.-W. and Pucci, G. and Silver, E. and Hooshanginejad, A. and Harris, D. M.},
  title   = {Synchronization of wave-propelled capillary spinners},
  journal = {Phys. Rev. E},
  volume  = {111},
  number  = {3},
  pages   = {035105},
  year    = {2025},
}

@article{womersley1955method,
  author  = {Womersley, J. R.},
  title   = {Method for the calculation of velocity, rate of flow and viscous drag in arteries when the pressure gradient is known},
  journal = {J. Physiol.},
  volume  = {127},
  number  = {3},
  pages   = {553},
  year    = {1955},
}

@book{schlichting2016boundary,
  author    = {Schlichting, H. and Gersten, K.},
  title     = {Boundary-Layer Theory},
  year      = {2016},
  publisher = {Springer},
}

@book{griffiths2023introduction,
  author    = {Griffiths, D. J.},
  title     = {Introduction to Electrodynamics},
  year      = {2023},
  publisher = {Cambridge University Press},
}

@article{anderson2005ludwig,
  author  = {Anderson, J. D.},
  title   = {Ludwig {Prandtl's} boundary layer},
  journal = {Phys. Today},
  volume  = {58},
  number  = {12},
  pages   = {42--48},
  year    = {2005},
}

@article{mukundarajan2016surface,
  author  = {Mukundarajan, H. and Bardon, T. C. and Kim, D. H. and Prakash, M.},
  title   = {Surface tension dominates insect flight on fluid interfaces},
  journal = {J. Exp. Biol.},
  volume  = {219},
  number  = {5},
  pages   = {752--766},
  year    = {2016},
}

@article{bush2006walking,
  author  = {Bush, J. W. M. and Hu, D. L.},
  title   = {Walking on water: Biolocomotion at the interface},
  journal = {Annu. Rev. Fluid Mech.},
  volume  = {38},
  number  = {1},
  pages   = {339--369},
  year    = {2006},
}

@article{hu2010hydrodynamics,
  author  = {Hu, D. L. and Bush, J. W. M.},
  title   = {The hydrodynamics of water-walking arthropods},
  journal = {J. Fluid Mech.},
  volume  = {644},
  pages   = {5--33},
  year    = {2010},
}

@article{suter1997locomotion,
  author  = {Suter, R. B. and Rosenberg, O. and Loeb, S. and Wildman, H. and Long, Jr., J. H.},
  title   = {Locomotion on the water surface: Propulsive mechanisms of the fisher spider {Dolomedes triton}},
  journal = {J. Exp. Biol.},
  volume  = {200},
  number  = {19},
  pages   = {2523--2538},
  year    = {1997},
}

@article{hsieh2004running,
  author  = {Hsieh, S. T. and Lauder, G. V.},
  title   = {Running on water: Three-dimensional force generation by basilisk lizards},
  journal = {Proc. Natl. Acad. Sci. U.S.A.},
  volume  = {101},
  number  = {48},
  pages   = {16784--16788},
  year    = {2004},
}

@article{rhee2022surferbot,
  author  = {Rhee, E. and Hunt, R. and Thomson, S. J. and Harris, D. M.},
  title   = {{SurferBot}: A wave-propelled aquatic vibrobot},
  journal = {Bioinspir. Biomim.},
  volume  = {17},
  number  = {5},
  pages   = {055001},
  year    = {2022},
}

@article{chen2018controllable,
  author  = {Chen, Y. and Doshi, N. and Goldberg, B. and Wang, H. and Wood, R. J.},
  title   = {Controllable water surface to underwater transition through electrowetting in a hybrid terrestrial-aquatic microrobot},
  journal = {Nat. Commun.},
  volume  = {9},
  number  = {1},
  pages   = {2495},
  year    = {2018},
}

@article{song2007surface,
  author  = {Song, Y. S. and Sitti, M.},
  title   = {Surface-tension-driven biologically inspired water strider robots: Theory and experiments},
  journal = {IEEE Trans. Robot.},
  volume  = {23},
  number  = {3},
  pages   = {578--589},
  year    = {2007},
}

@article{izri2014self,
  author  = {Izri, Z. and van der Linden, M. N. and Michelin, S. and Dauchot, O.},
  title   = {Self-propulsion of pure water droplets by spontaneous {Marangoni}-stress-driven motion},
  journal = {Phys. Rev. Lett.},
  volume  = {113},
  number  = {24},
  pages   = {248302},
  year    = {2014},
}

@article{bormashenko2015self,
  author  = {Bormashenko, E. and Bormashenko, Y. and Grynyov, R. and Aharoni, H. and Whyman, G. and Binks, B. P.},
  title   = {Self-propulsion of liquid marbles: {Leidenfrost}-like levitation driven by {Marangoni} flow},
  journal = {J. Phys. Chem. C},
  volume  = {119},
  number  = {18},
  pages   = {9910--9915},
  year    = {2015},
}

@book{faltinsen2005hydrodynamics,
  author    = {Faltinsen, O. M.},
  title     = {Hydrodynamics of High-Speed Marine Vehicles},
  year      = {2005},
  publisher = {Cambridge University Press},
}

@article{hunt2023drag,
  author  = {Hunt, R. and Zhao, Z. and Silver, E. and Yan, J. and Bazilevs, Y. and Harris, D. M.},
  title   = {Drag on a partially immersed sphere at the capillary scale},
  journal = {Phys. Rev. Fluids},
  volume  = {8},
  number  = {8},
  pages   = {084003},
  year    = {2023},
}

@article{zhou2022drag,
  author  = {Zhou, Z. and Vlahovska, P. M. and Miksis, M. J.},
  title   = {Drag force on spherical particles trapped at a liquid interface},
  journal = {Phys. Rev. Fluids},
  volume  = {7},
  number  = {12},
  pages   = {124001},
  year    = {2022},
}

@article{dorr2016drag,
  author  = {D{\"o}rr, A. and Hardt, S. and Masoud, H. and Stone, H. A.},
  title   = {Drag and diffusion coefficients of a spherical particle attached to a fluid--fluid interface},
  journal = {J. Fluid Mech.},
  volume  = {790},
  pages   = {607--618},
  year    = {2016},
}

@article{petkov1995measurement,
  author  = {Petkov, J. T. and Denkov, N. D. and Danov, K. D. and Velev, O. D. and Aust, R. and Durst, F.},
  title   = {Measurement of the drag coefficient of spherical particles attached to fluid interfaces},
  journal = {J. Colloid Interface Sci.},
  volume  = {172},
  number  = {1},
  pages   = {147--154},
  year    = {1995},
}

@article{ally2010magnetophoretic,
  author  = {Ally, J. and Amirfazli, A.},
  title   = {Magnetophoretic measurement of the drag force on partially immersed microparticles at air--liquid interfaces},
  journal = {Colloids Surf. A: Physicochem. Eng. Asp.},
  volume  = {360},
  number  = {1--3},
  pages   = {120--128},
  year    = {2010},
}

@article{barratt1965wave,
  author  = {Barratt, M. J.},
  title   = {The wave drag of a hovercraft},
  journal = {J. Fluid Mech.},
  volume  = {22},
  number  = {1},
  pages   = {39--47},
  year    = {1965},
}

@article{raphael1996capillary,
  author  = {Rapha{\"e}l, E. and de Gennes, P.-G.},
  title   = {Capillary gravity waves caused by a moving disturbance: Wave resistance},
  journal = {Phys. Rev. E},
  volume  = {53},
  number  = {4},
  pages   = {3448},
  year    = {1996},
}

@article{le2011wave,
  author  = {Le Merrer, M. and Clanet, C. and Qu{\'e}r{\'e}, D. and Rapha{\"e}l, E. and Chevy, F.},
  title   = {Wave drag on floating bodies},
  journal = {Proc. Natl. Acad. Sci. U.S.A.},
  volume  = {108},
  number  = {37},
  pages   = {15064--15068},
  year    = {2011},
}

@article{closa2010capillary,
  author  = {Closa, F. and Chepelianskii, A. D. and Rapha{\"e}l, E.},
  title   = {Capillary-gravity waves generated by a sudden object motion},
  journal = {Phys. Fluids},
  volume  = {22},
  number  = {5},
  pages   = {052107},
  year    = {2010},
}

@article{loudet2020drag,
  author  = {Loudet, J.-C. and Qiu, M. and Hemauer, J. and Feng, J. J.},
  title   = {Drag force on a particle straddling a fluid interface: Influence of interfacial deformations},
  journal = {Eur. Phys. J. E},
  volume  = {43},
  number  = {2},
  pages   = {13},
  year    = {2020},
}

@book{stokes1922mathematical,
  author    = {Stokes, G. G.},
  title     = {Mathematical and Physical Papers},
  volume    = {3},
  year      = {1922},
  publisher = {Cambridge University Press},
}

@article{rayleigh1911lxxxii,
  author  = {Rayleigh, Lord},
  title   = {{LXXXII}. On the motion of solid bodies through viscous liquid},
  journal = {Philos. Mag.},
  volume  = {21},
  number  = {126},
  pages   = {697--711},
  year    = {1911},
}

@article{panton1968transient,
  author  = {Panton, R.},
  title   = {The transient for {Stokes's} oscillating plate: A solution in terms of tabulated functions},
  journal = {J. Fluid Mech.},
  volume  = {31},
  number  = {4},
  pages   = {819--825},
  year    = {1968},
}

@article{liu2006note,
  author  = {Liu, C.-M. and Liu, I.-C.},
  title   = {A note on the transient solution of {Stokes'} second problem with arbitrary initial phase},
  journal = {J. Mech.},
  volume  = {22},
  number  = {4},
  pages   = {349--354},
  year    = {2006},
}

@article{erdogan2000note,
  author  = {Erdogan, M. E.},
  title   = {A note on an unsteady flow of a viscous fluid due to an oscillating plane wall},
  journal = {Int. J. Non-Linear Mech.},
  volume  = {35},
  number  = {1},
  pages   = {1--6},
  year    = {2000},
}

@article{li1971stokes,
  author  = {Li, K. W. and Marfatia, A. C.},
  title   = {{Stokes} second problem for the cylinder},
  journal = {J. Basic Eng.},
  volume  = {93},
  pages   = {326--328},
  year    = {1971},
}

@article{rivero2019study,
  author  = {Rivero, M. and Garz{\'o}n, F. and Nunez, J. and Figueroa, A.},
  title   = {Study of the flow induced by circular cylinder performing torsional oscillation},
  journal = {Eur. J. Mech. B/Fluids},
  volume  = {78},
  pages   = {245--251},
  year    = {2019},
}

@article{song2020viscous,
  author  = {Song, Y. and Rau, M. J.},
  title   = {Viscous fluid flow inside an oscillating cylinder and its extension to {Stokes'} second problem},
  journal = {Phys. Fluids},
  volume  = {32},
  number  = {4},
  pages   = {043603},
  year    = {2020},
}

@article{wang1965flow,
  author  = {Wang, C.-Y.},
  title   = {The flow field induced by an oscillating sphere},
  journal = {J. Sound Vib.},
  volume  = {2},
  number  = {3},
  pages   = {257--269},
  year    = {1965},
}

@article{riley1966sphere,
  author  = {Riley, N.},
  title   = {On a sphere oscillating in a viscous fluid},
  journal = {Q. J. Mech. Appl. Math.},
  volume  = {19},
  number  = {4},
  pages   = {461--472},
  year    = {1966},
}

@article{longuet1953mass,
  author  = {Longuet-Higgins, M. S.},
  title   = {Mass transport in water waves},
  journal = {Philos. Trans. R. Soc. Lond. A},
  volume  = {245},
  number  = {903},
  pages   = {535--581},
  year    = {1953},
}

@article{blennerhassett2002linear,
  author  = {Blennerhassett, P. J. and Bassom, A. P.},
  title   = {The linear stability of flat {Stokes} layers},
  journal = {J. Fluid Mech.},
  volume  = {464},
  pages   = {393--410},
  year    = {2002},
}

@article{von1974linear,
  author  = {Von Kerczek, C. and Davis, S. H.},
  title   = {Linear stability theory of oscillatory {Stokes} layers},
  journal = {J. Fluid Mech.},
  volume  = {62},
  number  = {4},
  pages   = {753--773},
  year    = {1974},
}

@article{hall1978linear,
  author  = {Hall, P.},
  title   = {The linear stability of flat {Stokes} layers},
  journal = {Proc. R. Soc. Lond. A},
  volume  = {359},
  number  = {1697},
  pages   = {151--166},
  year    = {1978},
}

@article{ai2005investigation,
  author  = {Ai, L. and Vafai, K.},
  title   = {An investigation of {Stokes'} second problem for non-{Newtonian} fluids},
  journal = {Numer. Heat Transf. A},
  volume  = {47},
  number  = {10},
  pages   = {955--980},
  year    = {2005},
}

@article{fetecau2009note,
  author  = {Fetecau, Corina and Jamil, M. and Fetecau, Constantin and Siddique, I.},
  title   = {A note on the second problem of {Stokes} for {Maxwell} fluids},
  journal = {Int. J. Non-Linear Mech.},
  volume  = {44},
  number  = {10},
  pages   = {1085--1090},
  year    = {2009},
}

@article{ishfaq2019stokes,
  author  = {Ishfaq, N. and Khan, W. A. and Khan, Z. H.},
  title   = {The {Stokes'} second problem for nanofluids},
  journal = {J. King Saud Univ. Sci.},
  volume  = {31},
  number  = {1},
  pages   = {61--65},
  year    = {2019},
}

@article{pal2015fluid,
  author  = {Pal, D. and Chakraborty, S.},
  title   = {Fluid flow induced by periodic temperature oscillation over a flat plate: Comparisons with the classical {Stokes} problems},
  journal = {Phys. Fluids},
  volume  = {27},
  number  = {5},
  pages   = {053602},
  year    = {2015},
}

@article{hehner2019stokes,
  author  = {Hehner, M. T. and Gatti, D. and Kriegseis, J.},
  title   = {{Stokes}-layer formation under absence of moving parts---A novel oscillatory plasma actuator design for turbulent drag reduction},
  journal = {Phys. Fluids},
  volume  = {31},
  number  = {5},
  pages   = {051701},
  year    = {2019},
}

@article{fitzgibbon2014scaling,
  author  = {Fitzgibbon, S. and Shaqfeh, E. S. G. and Fuller, G. G. and Walker, T. W.},
  title   = {Scaling analysis and mathematical theory of the interfacial stress rheometer},
  journal = {J. Rheol.},
  volume  = {58},
  number  = {4},
  pages   = {999--1038},
  year    = {2014},
}

@article{braunreuther2023nondestructive,
  author  = {Braunreuther, M. and Liegeois, M. and Fahy, J. V. and Fuller, G. G.},
  title   = {Nondestructive rheological measurements of biomaterials with a magnetic microwire rheometer},
  journal = {J. Rheol.},
  volume  = {67},
  number  = {2},
  pages   = {579--588},
  year    = {2023},
}

@article{shih2024viscoelastic,
  author  = {Shih, A. and Chung, S. J. and Shende, O. B. and Herwald, S. E. and Vezeridis, A. M. and Fuller, G. G.},
  title   = {Viscoelastic measurements of abscess fluids using a magnetic stress rheometer},
  journal = {Phys. Fluids},
  volume  = {36},
  number  = {11},
  pages   = {113603},
  year    = {2024},
}

@article{zell2016linear,
  author  = {Zell, Z. A. and Mansard, V. and Wright, J. and Kim, K. and Choi, S. Q. and Squires, T. M.},
  title   = {Linear and nonlinear microrheometry of small samples and interfaces using microfabricated probes},
  journal = {J. Rheol.},
  volume  = {60},
  number  = {1},
  pages   = {141--159},
  year    = {2016},
}

@article{marangoni1972principle,
  author  = {Marangoni, C.},
  title   = {The principle of the surface viscosity of liquids established by {Mr. J. Plateau}},
  journal = {Nuovo Cim.},
  volume  = {5},
  pages   = {239--273},
  year    = {1872},
}

@book{plateau1873statique,
  author    = {Plateau, J.},
  title     = {Statique Exp{\'e}rimentale et Th{\'e}orique des Liquides Soumis aux Seules Forces Mol{\'e}culaires},
  volume    = {2},
  year      = {1873},
  publisher = {Gauthier-Villars},
}

@article{manikantan2020surfactant,
  author  = {Manikantan, H. and Squires, T. M.},
  title   = {Surfactant dynamics: Hidden variables controlling fluid flows},
  journal = {J. Fluid Mech.},
  volume  = {892},
  pages   = {P1},
  year    = {2020},
}

@article{foffano2012colloids,
  author  = {Foffano, G. and Lintuvuori, J. S. and Stratford, K. and Cates, M. E. and Marenduzzo, D.},
  title   = {Colloids in active fluids: Anomalous microrheology and negative drag},
  journal = {Phys. Rev. Lett.},
  volume  = {109},
  number  = {2},
  pages   = {028103},
  year    = {2012},
}

@article{soni2019odd,
  author  = {Soni, V. and Bililign, E. S. and Magkiriadou, S. and Sacanna, S. and Bartolo, D. and Shelley, M. J. and Irvine, W. T. M.},
  title   = {The odd free surface flows of a colloidal chiral fluid},
  journal = {Nat. Phys.},
  volume  = {15},
  number  = {11},
  pages   = {1188--1194},
  year    = {2019},
}

@article{ruhs2013situ,
  author  = {R{\"u}hs, P. A. and B{\"o}ni, L. and Fuller, G. G. and Inglis, R. F. and Fischer, P.},
  title   = {In-situ quantification of the interfacial rheological response of bacterial biofilms to environmental stimuli},
  journal = {PLoS One},
  volume  = {8},
  number  = {11},
  pages   = {e78524},
  year    = {2013},
}

@article{knevzevic2021oscillatory,
  author  = {Kne{\v{z}}evi{\'c}, M. and Avil{\'e}s Podgurski, L. E. and Stark, H.},
  title   = {Oscillatory active microrheology of active suspensions},
  journal = {Sci. Rep.},
  volume  = {11},
  number  = {1},
  pages   = {22706},
  year    = {2021},
}

@article{miller1968oscillations,
  title={The oscillations of a fluid droplet immersed in another fluid},
  author={Miller, CA and Scriven, LE},
  journal={J. Fluid Mech.},
  volume={32},
  number={3},
  pages={417--435},
  year={1968},
  publisher={Cambridge University Press}
}

@article{kamoliddinov2021hydrodynamic,
  title={Hydrodynamic regimes and drag on horizontally pulled floating spheres},
  author={Kamoliddinov, F and Vakarelski, IU and Thoroddsen, ST},
  journal={Phys. Fluids},
  volume={33},
  number={9},
  pages={093308},
  year={2021},
  publisher={AIP Publishing}
}

@book{newman1977,
  author    = {Newman, J. N.},
  title     = {Marine Hydrodynamics},
  publisher = {MIT Press},
  address   = {Cambridge, MA},
  year      = {1977},
}

@article{morison1950,
  author  = {Morison, J. R. and O'Brien, M. P. and Johnson, J. W. and Schaaf, S. A.},
  title   = {The force exerted by surface waves on piles},
  journal = {J. Pet. Technol.},
  volume  = {2},
  number  = {5},
  pages   = {149--154},
  year    = {1950},
}

@book{sarpkaya2010,
  author    = {Sarpkaya, T.},
  title     = {Wave Forces on Offshore Structures},
  publisher = {Cambridge University Press},
  address   = {Cambridge, UK},
  year      = {2010},
}

@article{danov1995influence,
  author  = {Danov, K. D. and Aust, R. and Durst, F. and Lange, U.},
  title   = {Influence of the surface viscosity on the hydrodynamic resistance and surface diffusivity of a large {Brownian} particle},
  journal = {J. Colloid Interface Sci.},
  volume  = {175},
  number  = {1},
  pages   = {36--45},
  year    = {1995},
}

@article{danov2000viscous,
  author  = {Danov, K. D. and Dimova, R. and Pouligny, B.},
  title   = {Viscous drag of a solid sphere straddling a spherical or flat surface},
  journal = {Phys. Fluids},
  volume  = {12},
  number  = {11},
  pages   = {2711--2722},
  year    = {2000},
}

@article{pozrikidis2007particle,
  author  = {Pozrikidis, C.},
  title   = {Particle motion near and inside an interface},
  journal = {J. Fluid Mech.},
  volume  = {575},
  pages   = {333--357},
  year    = {2007},
}

@article{dani2015hydrodynamics,
  author  = {Dani, A. and Keiser, G. and Yeganeh, M. and Maldarelli, C.},
  title   = {Hydrodynamics of particles at an oil--water interface},
  journal = {Langmuir},
  volume  = {31},
  number  = {49},
  pages   = {13290--13302},
  year    = {2015},
}

@article{ackerberg1972unsteady,
  author  = {Ackerberg, R. C. and Phillips, J. H.},
  title   = {The unsteady laminar boundary layer on a semi-infinite flat plate due to small fluctuations in the magnitude of the free-stream velocity},
  journal = {J. Fluid Mech.},
  volume  = {51},
  number  = {1},
  pages   = {137--157},
  year    = {1972},
}

@article{rott1960response,
  title={On the response of the laminar boundary layer to small fluctuations of the free-stream velocity},
  author={Rott, Nicholas and Rosenzweig, Martin L},
  journal={J. Aeronaut. Sci.},
  volume={27},
  number={10},
  pages={741--747},
  year={1960}
}

@article{lighthill1954response,
  title={The response of laminar skin friction and heat transfer to fluctuations in the stream velocity},
  author={Lighthill, Michael James},
  journal={Proc. R. Soc. Lond. Ser. A-Math. Phys. Sci.},
  volume={224},
  number={1156},
  pages={1--23},
  year={1954},
  publisher={The Royal Society London}
}

@article{leonard2001aspects,
  title={Aspects of flow-induced vibration},
  author={Leonard, A and Roshko, A},
  journal={Journal of Fluids and Structures},
  volume={15},
  number={3-4},
  pages={415--425},
  year={2001},
  publisher={Elsevier}
}

@article{chepelianskii2008capillary,
  title={Capillary-gravity waves generated by a slow moving object},
  author={Chepelianskii, AD and Chevy, Fr{\'e}d{\'e}ric and Raphael, Elie},
  journal={Phys. Rev. Lett.},
  volume={100},
  number={7},
  pages={074504},
  year={2008},
  publisher={APS}
}

\end{document}